\documentclass{aa}

\usepackage{graphicx}

\usepackage{txfonts}

\usepackage[colorlinks=true,
    citecolor=blue]{hyperref}

\newcommand{\rhoi}{\rho_{\mathrm{i}}}
\newcommand{\rhoh}{\rho_{\mathrm{H}}}
\newcommand{\rhohe}{\rho_{\mathrm{He}}}

\newcommand{\aih}{\alpha_{\rm i\,H}}
\newcommand{\aihe}{\alpha_{\rm i\,He}}
\newcommand{\ahhe}{\alpha_{\rm H\,He}}

\newcommand{\ahi}{\alpha_{\rm H\,i}}
\newcommand{\ahei}{\alpha_{\rm He\,i}}
\newcommand{\aheh}{\alpha_{\rm He\,H}}

\newcommand{\ahetot}{\alpha_{{\rm He}}}

\begin{document}

 %  \title{Torsional Alfv\'en Waves propagating through the lower solar atmosphere: Putting the  single-fluid approximation to the test}

 %  \title{Testing the single-fluid approximation for  Alfv\'en wave propagation in the lower solar atmosphere}

%\title{Torsional Alfv\'en wave propagation in the partially ionized lower solar atmosphere: a test of the single-fluid approximation}

\title{Alfv\'en wave propagation in the partially ionized lower solar atmosphere: a test of the single-fluid approximation}

\titlerunning{Testing the single-fluid approximation}
\authorrunning{R. Soler}

     \author{Roberto Soler\inst{1,2}}

   \institute{Departament de Física, Universitat de les Illes Balears, 07122 Palma de Mallorca, Spain\\
            \and Institut d’Aplicacions Computacionals de Codi Comunitari (IAC3), Universitat de les Illes Balears, 07122 Palma de Mallorca,
            Spain\\ 
             \email{roberto.soler@uib.es}}

   \date{Received XXX; accepted XXX}

  \abstract{

  Alfv\'en waves are widely believed to play an important role in the transport of energy from the solar photosphere to the corona through the partially ionized chromosphere. In previous work, the properties of torsional Alfv\'en waves were theoretically studied using a multi-fluid model. Here, we compare those multi-fluid results with those obtained using the single-fluid magnetohydrodynamic approximation, as a way to assess the performance of the latter in the context of Alfv\'enic waves in the lower solar atmosphere. We consider a broadband photospheric driver that excites torsional Alfv\'en waves with frequencies ranging from 0.1~mHz to 300~mHz. These waves propagate upwards to the corona along a magnetic flux tube expanding with height. For both models, we compare the energy flux, chromospheric reflection, transmission and absorption coefficients, and the associated heating rates. In general, the results are almost identical in the two models, with the exception of two minor differences: (1) the net energy flux reaching the corona is approximately 5\% larger in the single-fluid model, mainly owing to the higher reflectivity found in the multi-fluid model for wave frequencies exceeding 10~mHz; and (2) in a narrow region around 500~km above the photosphere, the single-fluid model underestimates the plasma heating rate due to ion-neutral damping by about a factor of two compared with the multi-fluid model. Both discrepancies arise from the approximate treatment of the ion-neutral drift in the single-fluid model and are expected to have a very limited impact on practical applications.}


   \maketitle

\section{Introduction}

Alfv\'en waves are a type of incompressible magnetohydrodynamic (MHD) wave whose excitation in the solar atmosphere by photospheric granulation was postulated by Alfv\'en himself \citep{alfven1947}. In cylindrically symmetric magnetic flux tubes, pure Alfv\'en modes\footnote{By pure Alfv\'en modes we mean Alfv\'en waves that are not coupled with another type of mode.} correspond to torsional waves, which produce axisymmetric velocity and magnetic field perturbations polarized in the azimuthal direction \citep[see, e.g.,][]{erdelyi2007}. There is observational evidence of torsional Alfv\'en waves propagating through various layers of the solar atmosphere: the photosphere \citep{jess2009}, the chromosphere and transition region \citep{depontieu2012,depontieu2014}, and the corona \citep{morton2026}. It has been theoretically shown that these waves can partially overcome the strong filtering caused by chromospheric reflection and dissipation, and thus provide a significant energy input to the coronal medium \citep[see, e.g.,][]{soler2017,soler2019}. In coronal flux tubes, these torsional waves can nonlinearly generate turbulence \citep[see, e.g.,][]{guo2019,diaz2021}.

The solar chromosphere is a partially ionized region \citep[see, e.g.,][]{sykora2015}, and it is well known that partial ionization of the plasma strongly affects the properties of MHD waves \citep[see the review by][]{soler2024}. The theoretical study of MHD waves in partially ionized solar plasma is usually carried out using two approaches: single-fluid and multi-fluid models \citep[see][]{khomenko2014,ballester2018}. The single-fluid approximation assumes that all species (in particular, ions and neutrals) are strongly coupled, and the effects of their interaction appear as nonideal terms in the governing equations. Among these nonideal terms, ambipolar diffusion plays a predominant role. Some representative wave studies in the solar context that use the single-fluid approximation include, for example, \citet{Khodachenko2004,Leake2005,forteza2007,goodman2011,arber2016,cally2019}. Conversely, in the multi-fluid description the degree of coupling between different species is allowed to be arbitrary, and interaction terms are explicitly included in the equations. A particular version of the multi-fluid description is the two-fluid model, in which charged particles and neutrals form two separate fluids. Examples of wave studies using the multi-fluid approach include \citet{zaqarashvili2011,song2011,tu2013,maneva2017,martinez2017,martinez2018,popescu2019,Pelekhata2021,cally2023}, among many others.

The multi-fluid model provides a more accurate physical description than the single-fluid approximation, although its use in analytical studies and its implementation in numerical codes are both more involved. Nevertheless, owing to the high values of collision frequencies in the lower solar atmosphere \citep[see][]{ballester2018}, the single-fluid approximation often yields a sufficiently accurate description, although its accuracy may depend on the specific process under investigation. The purpose of this paper is to revisit the study of torsional Alfv\'en wave propagation from the photosphere to the corona presented in \citet{soler2019}. In that earlier work, a multi-fluid model was employed. Here, we aim to compare those multi-fluid results with those obtained using the single-fluid model, as a means of assessing the performance of the single-fluid approximation in the context of Alfv\'en waves in the partially ionized lower solar atmosphere. Recently, \citet{gomez2025} carried out an investigation with a similar goal, although their study and the present work differ in several aspects, such as the background configuration, methodology, and the type of wave studied.

This paper is organized as follows. Section~\ref{sec:model} describes the model, the governing equations, and the methodology. Section~\ref{sec:res} presents a comparison between the multi-fluid and single-fluid results. Finally, Section~\ref{sec:conc} provides a discussion and concluding remarks.

\section{Model, basic equations, and method}
\label{sec:model}

\subsection{Background atmosphere}

The background atmosphere and magnetic field model are the same as those used in \citet{soler2019}. We refer readers to the previous paper for detailed explanations; here, we provide only a brief summary.  

The atmosphere is a gravitationally stratified medium whose physical properties vary with height according to an adapted version of the quiet-Sun chromospheric model C of \citet{FAL93}. The atmosphere is plane-parallel, so the plasma properties are invariant in the horizontal directions. The model extends from the base of the photosphere through the chromosphere and the transition region, reaching the low corona at 4,000~km above the photosphere. The plasma is composed of hydrogen and helium and is partially ionized. A strong thermal coupling is assumed between all species, so they share a common temperature. The variation of the total density and temperature with height according to this model can be seen in Figure~1 of \citet{soler2019}.  

\begin{figure}
   \centering
\includegraphics[width=0.7\columnwidth]{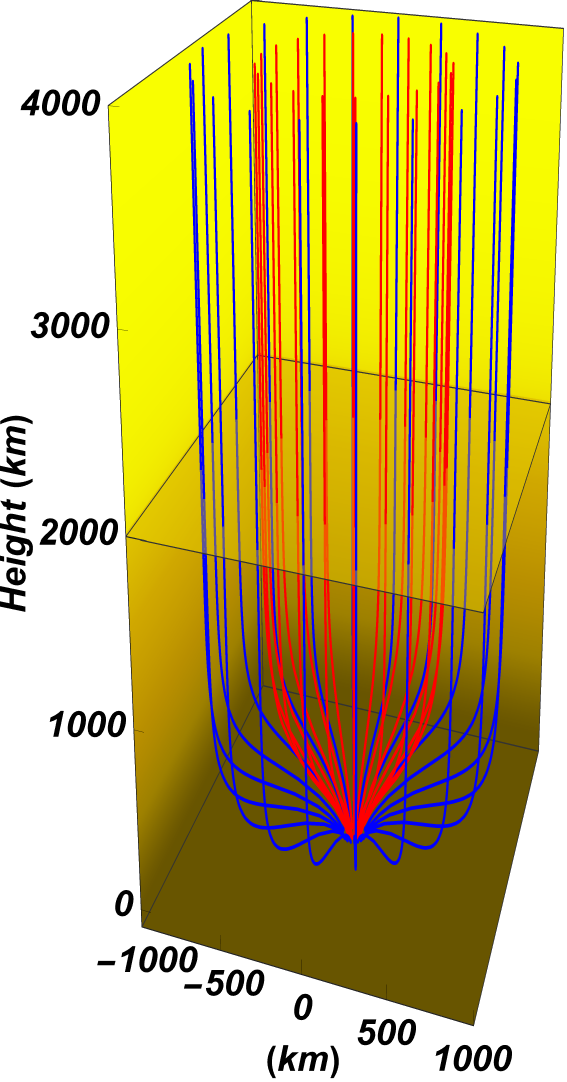} 
      \caption{Equilibrium magnetic flux tube model embedded in the background  atmosphere. The red and blue lines outline some selected magnetic field lines that cross the photosphere at 40~km and 70~km from the tube axis, respectively. The color gradient illustrates the density stratification from the photosphere (brown) to the low corona (yellow). The transition region height  is marked by a horizontal, semi-transparent plane. For visualization purposes, the horizontal and vertical directions are not to scale.}
         \label{fig:tube}
   \end{figure}

The magnetic field configuration consists of a potential flux tube that expands with height from the photosphere to the corona. The tube axis is vertical, and the field lines are untwisted. We use cylindrical coordinates with the $z$-axis directed along the flux tube axis, which coincides with the vertical direction. Hence, the background magnetic field vector, $\mathbf{B}$, is expressed as  
\begin{equation}
\mathbf{B} = B_r \left( r, z \right) \hat{e}_r + B_z \left( r, z \right) \hat{e}_z.
\end{equation}  
The magnetic flux tube is numerically constructed by imposing that $\mathbf{B}$ satisfies both the current-free ($\nabla \times \mathbf{B} = 0$) and divergence-free ($\nabla \cdot \mathbf{B} = 0$) conditions. At the photosphere, the tube radius is set to 100~km, with a magnetic field strength at the tube center of 1~kG. The radius increases with height due to tube expansion, reaching 1,000~km in the low corona. Because of conservation of magnetic flux, the magnetic field strength decreases with height, reaching a minimum of 10~G in the low corona, where the field becomes purely vertical. A three-dimensional visualization of the flux tube embedded in the stratified atmosphere can be seen in Figure~\ref{fig:tube}.

\subsection{From multi-fluid equations to the single-fluid approximation}

The propagation of linear torsional Alfv\'en waves from the photosphere to the corona in the model described above was investigated in \citet{soler2019} using a multi-fluid treatment. All ions (protons, singly ionized helium, and doubly ionized helium) were treated together as a single ionic fluid, while neutral hydrogen and neutral helium were treated as two separate neutral fluids. The three fluids exchanged momentum through particle collisions. Electron inertia was neglected, while collisions of electrons with the other species gave rise to the magnetic diffusion (i.e., resistivity) term in the induction equation.  

Here, the goal is to use the single-fluid approximation to further reduce the number of independent fluids considered in the plasma treatment. To this end, we closely follow the derivation given in \citet{zaqarashvili2013}. Readers are also referred to \citet{khomenko2014}, \citet{ballester2018}, and references therein for more details on the physical and mathematical subtleties of this approach.

\begin{figure}
   \centering
\includegraphics[width=0.99\columnwidth]{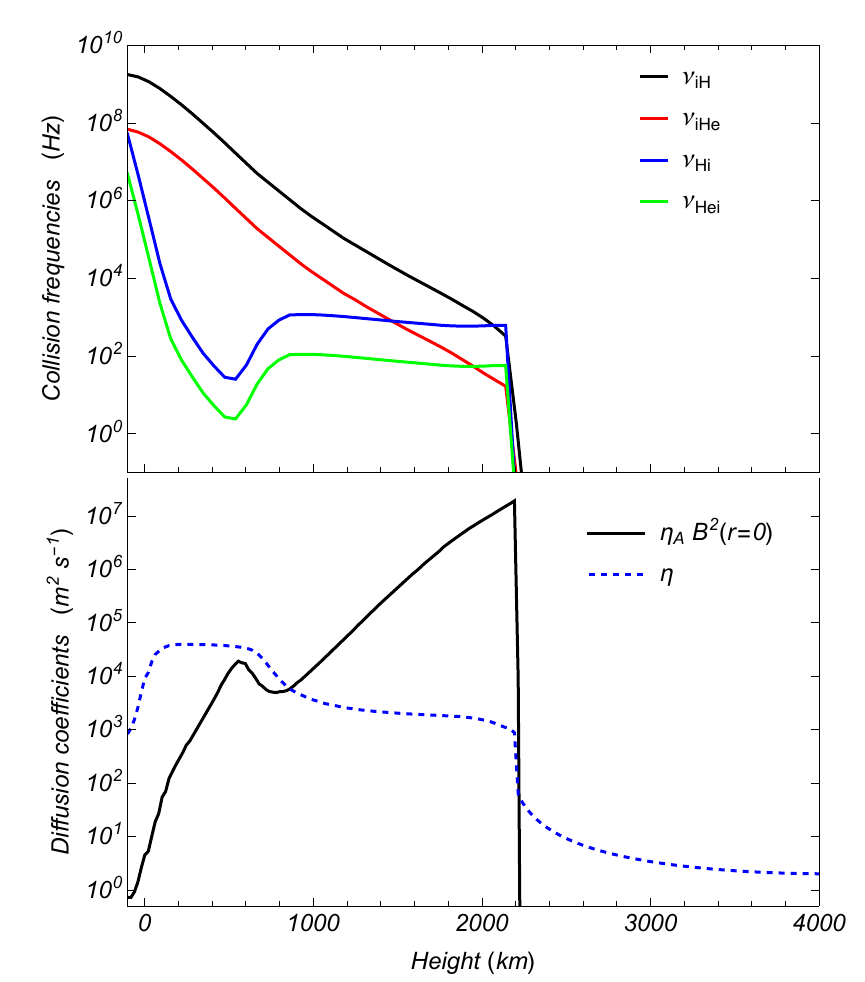} 
      \caption{Dependence on height above the photosphere of the  collision frequencies involving neutrals (top) and the Ohmic and ambipolar diffusion coefficients (bottom) according to the assumed background atmospheric model. The  ambipolar coefficient displayed in this plot has been computed using the magnetic field strength at the axis of the flux tube.}
         \label{fig:model}
   \end{figure}

 We start from the linearized equations of motion and the linearized induction equation used in  \citet{soler2019}, namely
\begin{eqnarray}
\rhoi \frac{\partial {\bf v'_{\rm i}}}{\partial t} &=& \frac{1}{\mu} \left( \nabla \times {\bf B'} \right)\times {\bf B} - \aih \left( {\bf v'_{\rm i}} - {\bf v'_{\rm H}}  \right)  \nonumber \\ &&  - \aihe \left( {\bf v'_{\rm i}} - {\bf v'_{\rm He}}  \right), \label{eq:momi} \\
\rhoh \frac{\partial {\bf v'_{\rm H}}}{\partial t} &=&  - \ahi \left( {\bf v'_{\rm H}} - {\bf v'_{\rm i}}  \right)     - \ahhe \left( {\bf v'_{\rm H}} - {\bf v'_{\rm He}}  \right), \label{eq:momh} \\
\rhohe \frac{\partial {\bf v'_{\rm He}}}{\partial t} &=&  - \ahei \left( {\bf v'_{\rm He}} - {\bf v'_{\rm i}}  \right)     - \aheh \left( {\bf v'_{\rm He}} - {\bf v'_{\rm H}}  \right), \label{eq:momhe} \\
\frac{\partial \bf B'}{\partial t} &=& \nabla \times \left( {\bf v'_{\rm i}} \times {\bf B} \right) - \nabla \times \left[ \eta   \left( \nabla \times {\bf B'} \right) \right]. \label{eq:induction}
\end{eqnarray}
In these equations, subscripts i, H, and He denote the ions, neutral hydrogen, and neutral helium, respectively. We shall use subscript $\beta$ to refer to an arbitrary species, so that $\rho_\beta$ and ${\bf v'_{\beta}}$ are the mass density and velocity perturbation of species $\beta$, respectively. In turn, $\bf B$ and $\bf B'$ are the background  magnetic field (defined before) and the magnetic field  perturbation, respectively. Finally, $\alpha_{\beta\beta'} = \alpha_{\beta'\beta}$ is the symmetric friction coefficient for collisions between species $\beta$ and $\beta'$,  and $\eta$ is the coefficient of resistivity or Ohmic diffusion. The expressions of these various coefficients are given \citet{soler2019} but are  omitted here for the sake of simplicity. The dependence of $\eta$ on height in the atmospheric model is shown in the bottom panel of Figure~\ref{fig:model}. We note that gas pressure terms are omitted from these equations because they are irrelevant for the investigation of Alfv\'en waves.

The process of deriving the single-fluid equations begins by defining the total density, $\rho$, and the center-of-mass velocity perturbation, $\bf v'$, as
\begin{eqnarray}
\rho & \equiv & \rhoi + \rhoh + \rhohe, \\ 
{\bf v'} & \equiv & \xi_{\rm i} {\bf v'_{\rm i}} + \xi_{\rm H}  {\bf v'_{\rm H}} + \xi_{{\rm He}}  {\bf v'_{\rm He}}, \label{eq:vcm}
\end{eqnarray}
where $\xi_\beta = \rho_\beta/\rho$ is the fraction of species $\beta$. To obtain approximate expressions for the ion-neutral velocity drifts, we perform consecutive subtractions of the various momentum equations and neglect the differences of the inertia terms \citep[see details in][]{zaqarashvili2013}. The velocity drifts are found to be proportional to the Lorentz force and  can be cast as,
\begin{eqnarray}
{\bf v'_{\rm i}} - {\bf v'_{\rm H}} & \approx & \mathcal{C}_{\rm H} \frac{1}{\mu}  \left( \nabla \times {\bf B'} \right)\times {\bf B},  \label{eq:driftih} \\
{\bf v'_{\rm i}} - {\bf v'_{\rm He}} & \approx & \mathcal{C}_{\rm He}    \frac{1}{\mu} \left( \nabla \times {\bf B'} \right)\times {\bf B}.   \label{eq:driftihe}
\end{eqnarray}
with the proportionality factors given by,
\begin{eqnarray}
\mathcal{C}_{\rm H} &=& \frac{\xi_{\rm H}\aihe +  \left(\xi_{\rm H} +\xi_{{\rm He}} \right)\ahhe}{\aih \aihe + \left( \aih  + \aihe\right)\ahhe}, \label{eq:ch} \\
\mathcal{C}_{\rm He} &=&  \frac{\xi_{{\rm He}}\aih +  \left(\xi_{\rm H} + \xi_{{\rm He}} \right)\ahhe}{\aih \aihe + \left( \aih  + \aihe\right)\ahhe}. \label{eq:che}
\end{eqnarray}
The validity of the approximate Equations~(\ref{eq:driftih}) and (\ref{eq:driftihe}) relies on the conditions that the wave frequency is much lower than the collision frequencies between ions and neutrals and that the magnitude of the velocity drifts are much smaller than the thermal velocity\footnote{A more general formulation has been discussed by \citet{hillier2024}.}. The frequency of collisions  between species $\beta$ and $\beta'$ is
\begin{equation}
    \nu_{\beta\beta'} = \frac{\alpha_{\beta\beta'}}{\rho_\beta}.
\end{equation}
The top panel of Figure~\ref{fig:model} shows the height dependence of the collision frequencies involving neutrals in the atmospheric model. We note that all these frequencies vanish above the transition region, where the plasma becomes fully ionized. The most restrictive frequency is that of collisions between neutral helium and ions, $\nu_{\rm He\,i}$, whose minimum value is approximately 2~Hz at 500~km above the photosphere. Therefore, waves with lower frequencies should satisfy the conditions for the applicability of the single-fluid approximation. We note that if the plasma were composed only of hydrogen, the most restrictive frequency would be that of collisions between neutral hydrogen and ions, $\nu_{\rm H\,i}$, which is approximately an order of magnitude larger than $\nu_{\rm He\,i}$ at those heights. The presence of helium reduces the range of applicability of the single-fluid approximation to lower frequencies. 

The relation between the ion-neutral drift associated with hydrogen and that associated with helium can be estimated by taking the ratio of Equations~(\ref{eq:driftih}) and (\ref{eq:driftihe}), namely
\begin{equation}
    \frac{{\bf v'_{\rm i}} - {\bf v'_{\rm H}}}{{\bf v'_{\rm i}} - {\bf v'_{\rm He}}} \approx \frac{\nu_{\rm He\,i}}{\nu_{\rm H\,i}},
\end{equation}
where, for simplicity,  collisions between the two neutral species have been neglected in this comparison. Since $\nu_{\rm He\,i} < \nu_{\rm H\,i}$ throughout the chromosphere, the ion-neutral drift associated with helium is expected to be larger than that associated with hydrogen.

By combining Equations~(\ref{eq:driftih}) and (\ref{eq:driftihe}) with Equation~(\ref{eq:vcm}), we can approximate the ion velocity perturbation as,
\begin{equation}
{\bf v'_{\rm i}} \approx {\bf v'} + \eta_{\rm A } \left( \nabla \times {\bf B'} \right)\times {\bf B}, \label{eq:vi}
\end{equation}
where the second term on the right-hand side represents the ambipolar drift, with $\eta_{\rm A}$ being the ambipolar coefficient, given by  
\begin{equation}
 \eta_{\rm A } = \frac{\xi_{\rm H}\mathcal{C}_{\rm H} + \xi_{{\rm He}}\mathcal{C}_{\rm He}}{\mu}.
\end{equation}
Using the expressions for $\mathcal{C}_{\rm H}$ and $\mathcal{C}_{\rm He}$ from Equations~(\ref{eq:ch}) and (\ref{eq:che}), and after some algebra, the ambipolar coefficient can be rewritten as  
\begin{equation}
 \eta_{\rm A } = \frac{\xi_{\rm H}^2 \ahetot + \xi^2_{{\rm He}}\alpha_{\rm H}  + 2\xi_{\rm H}\xi_{{\rm He}} \alpha_{{\rm H\, He}}}{\mu \left( \alpha_{\rm H} \alpha_{{\rm He}} - \alpha^2_{{\rm H\, He}} \right)}, \label{eq:etaa}
\end{equation}
where $\alpha_\beta = \sum_{\beta' \neq \beta} \alpha_{\beta\beta'}$ is the total friction coefficient of species $\beta$. Equation~(\ref{eq:etaa}) agrees with the expression of $\eta_{\rm A}$ for a hydrogen-helium plasma previously obtained by \citet{zaqarashvili2013}.

A comparison of the values of $\eta$ and $\eta_{\rm A}$ in the atmospheric model is shown in the bottom panel of Figure~\ref{fig:model}. We note that the SI units of $\eta_{\rm A}$ are m$^2$~s$^{-1}$~T$^{-2}$, so that the quantity that actually corresponds to a diffusion coefficient is $\eta_{\rm A} | \mathbf{B} |^2$. Therefore, the efficiency of ambipolar diffusion depends on the magnetic field strength. Assuming a magnetic field strength equal to that at the center of the flux tube, we find that ambipolar diffusion dominates over Ohmic diffusion from approximately 1,000~km above the photosphere up to the transition region, where the plasma becomes fully ionized. A similar plot can be seen in Figure~1 of \citet{khomenko2012}, although for a different magnetic field configuration.

The single-fluid version of the linearized momentum equation is obtained by simply adding Equations~(\ref{eq:momi}), (\ref{eq:momh}), and (\ref{eq:momhe}), so that the collisional terms cancel out, giving  
\begin{equation}
\rho \frac{\partial{\bf v'}}{\partial t} = \frac{1}{\mu} \left( \nabla \times {\bf B'} \right)\times {\bf B}. \label{eq:momentumsf}
\end{equation}
For the linearized induction equation (Equation~(\ref{eq:induction})), we use Equation~(\ref{eq:vi}) to express the ion velocity perturbation in terms of the center-of-mass velocity and the ambipolar drift. By doing so, a term accounting explicitly for ambipolar diffusion appears in the induction equation:  
\begin{eqnarray}
\frac{\partial \bf B'}{\partial t} &=& \nabla \times \left( {\bf v'} \times {\bf B} \right) - \nabla \times \left[ \eta   \left( \nabla \times {\bf B'} \right) \right] \nonumber \\
& &+ \nabla \times \left\{  \eta_{\rm A } \left[\left( \nabla \times {\bf B'} \right)\times {\bf B}\right] \times  {\bf B} \right\}, \label{eq:inductionsf}
\end{eqnarray}
where the ambipolar term is the third term on the right-hand side.

In summary, Equations~(\ref{eq:momentumsf}) and (\ref{eq:inductionsf}) form the basic linearized equations for discussing Alfv\'en waves in the single-fluid approximation. As in the initial multi-fluid Equations~(\ref{eq:momi})–(\ref{eq:induction}), gas pressure terms are omitted. Since we have already considered  linearized  equations at the beginning of the derivation, Equations~(\ref{eq:momentumsf}) and (\ref{eq:inductionsf}) are not generally applicable beyond the context of linear Alfv\'en waves, but they suffice for our present goals. The complete nonlinear single-fluid equations can be found in, e.g., \citet{khomenko2014} and \citet{ballester2018}.

\subsection{Governing equations for torsional Alfv\'en waves}

From here on, we focus on purely torsional Alfv\'en waves. We use cylindrical coordinates and assume that the only nonzero perturbations are the azimuthal components of $\bf v'$ and $\bf B'$, namely
\begin{equation}
{\bf v'} =  v'_{\varphi} \hat{e}_\varphi, \qquad {\bf B'} =  B'_\varphi \hat{e}_\varphi.
\end{equation}
Furthermore,  we consider the stationary state of wave propagation\footnote{The validity of this approach has been assessed in \citet{soler2025} against the full temporal evolution.}. We express the temporal dependence of $v'_{\varphi}$ and  $B'_{\varphi}$ as $\exp\left(-i\omega t\right)$, where $\omega$ is the angular frequency of the waves. Thus, from Equation~(\ref{eq:momentumsf}) we can write $v'_{\varphi}$ in terms of $ B'_\varphi $ as
\begin{equation}
v'_{\varphi} = \frac{i}{\omega} \frac{1}{\mu \rho} \frac{1}{r}  {\bf B}\cdot \nabla \left( r B'_\varphi  \right). \label{eq:vfi}
\end{equation}
Using Equation~(\ref{eq:vfi}) in Equation~(\ref{eq:inductionsf}) to replace $v'_{\varphi}$, we obtain a partial differential equation involving $B'_\varphi$ alone:  
\begin{eqnarray}
&&i\omega B'_\varphi + r {\bf B} \cdot \nabla \left[ \frac{i}{\omega} \frac{1}{\mu\rho} \frac{1}{r^2} {\bf B} \cdot \nabla \left( r B'_\varphi \right) \right]    +  \eta \left(  \nabla^2 B'_\varphi - \frac{1}{r^2}  B'_\varphi\right) \nonumber \\
 && + \frac{\partial \eta}{\partial z} \frac{\partial B'_\varphi}{\partial z} 
   +  \eta_{\rm A }  \left\{ {\bf B} \cdot \nabla \left[ \frac{1}{r} {\bf B} \cdot \nabla \left( r B'_\varphi \right) \right] - \frac{B_r}{r^2}{\bf B} \cdot \nabla \left( r B'_\varphi \right)  \right\}   \nonumber \\
 && + \frac{\partial \eta_{\rm A } }{\partial z} \frac{B_z}{r}{\bf B} \cdot \nabla \left( r B'_\varphi \right) =0. \label{eq:main}
\end{eqnarray}
Equation~(\ref{eq:main}) is our main equation. It corresponds to the single-fluid version of Equation~(36) of \citet{soler2019}, which was derived in the multi-fluid case. Two differences arise between the single-fluid and multi-fluid equations, which are related to the different treatment of the ion-neutral (ambipolar) drift. In the present Equation~(\ref{eq:main}), there are terms proportional to $\eta_{\rm A}$ and $\partial \eta_{\rm A}/\partial z$, which explicitly account for the ambipolar drift as a diffusion mechanism for the magnetic field in the single-fluid approximation. Conversely, in the multi-fluid version of the equation, the effect of the ion-neutral drift is more subtly included in the second term through the effective density, $\rho_{\rm eff}$, which accounts for the fact that the inertia of the plasma in response to magnetic field oscillations depends on the strength of the coupling between ions and neutrals. The effective density is defined in Equation~(34) of \citet{soler2019} and reduces to the total density, $\rho$, in the single-fluid limit, where all species are assumed to be strongly coupled. Thus, the second term in Equation~(\ref{eq:main}) consistently involves $\rho$ instead of $\rho_{\rm eff}$ as in the multi-fluid version.

Using Equations~(\ref{eq:momentumsf}) and (\ref{eq:inductionsf}), we derive an equation for the evolution of the wave energy following the method of \citet{walker2005}. To retain the net energy contribution, we average the energy equation in time over a full period of the waves.  The resulting time-averaged energy equation is
\begin{equation}
\frac{\partial \left<U\right>}{\partial t} + \nabla \cdot {\bf \left<\Pi\right>} = - \left<H\right>, \label{eq:energycon}
\end{equation}
where $\left<U\right>$ is the time-averaged total  energy density, ${\bf \left<\Pi\right>}$ is the time-averaged  energy flux, and $ \left<H\right>$ is the time-averaged   dissipated energy, which  represents a heating input for the background plasma. The expressions for these quantities are:
\begin{eqnarray}
\left<U\right>  &=& \frac{1}{4} \rho {v'_{\varphi}} {v'_{\varphi}}^*  + \frac{1}{4\mu}{B'_\varphi}{B'_\varphi}^* , \label{eq:energydensity} \\
{\bf \left<\Pi\right>}  &=& -\frac{1}{2\mu} {\rm Re} \left( v'_{\varphi} {B'_\varphi}^* \right) {\bf B}  - \frac{\eta_{\rm A}}{2\mu}\frac{1}{r}{\bf B} \cdot \nabla \left( r B'_\varphi \right){B'_\varphi}^* {\bf B} \nonumber \\
&& - \frac{\eta}{2\mu} \left( \frac{1}{r} \frac{\partial \left( r B'_\varphi \right)}{\partial r}{B'_\varphi}^* \hat{e}_r + \frac{\partial B'_\varphi}{\partial z}{B'_\varphi}^* \hat{e}_z  \right), \label{eq:energyflux} \\
\left<H\right> &=& \frac{\eta}{2\mu} \left( \frac{1}{r^2} \frac{\partial \left( r B'_\varphi \right)}{\partial r} \frac{\partial \left( r  {B'_\varphi}^* \right)}{\partial r}  + \frac{\partial B'_\varphi}{\partial z} \frac{\partial {B'_\varphi}^*}{\partial z}   \right) \nonumber \\
&& +  \frac{\eta_{\rm A}}{2\mu}  \frac{1}{r^2}{\bf B} \cdot \nabla \left( r B'_\varphi \right){\bf B} \cdot \nabla \left( r {B'_\varphi}^*\right), \label{eq:heat}
\end{eqnarray}
where $*$ denotes the complex conjugate.

The expression for the energy flux (Equation~(\ref{eq:energyflux})) contains three terms. The first term corresponds to the ideal Alfv\'en wave energy flux and is aligned exclusively along the direction of the background magnetic field. The second and third terms are corrections due to ambipolar diffusion (second term) and Ohmic diffusion (third term). The ambipolar-related term is also aligned with the background magnetic field, whereas the Ohmic-related term has a component perpendicular to the background field. Nevertheless, this perpendicular component is orders of magnitude smaller than the field-aligned component \citep[see][]{soler2019}.  

The energy flux in the multi-fluid case corresponds to Equation~(39) of \citet{soler2019}. An explicit ambipolar-related term is absent from the multi-fluid expression, but the ideal term is written in terms of the ion velocity perturbation instead of the center-of-mass velocity. This means that the effect of ambipolar diffusion on the energy flux is indeed included in the multi-fluid formalism, although not as an explicit term. Conversely, the Ohmic-related term is also present in Equation~(39) of \citet{soler2019}, but it was neglected in their computations. Here, we retain the Ohmic-related term and will demonstrate that its contribution is indeed negligible.  

Concerning the expression for the heating rate (Equation~(\ref{eq:heat})), the term proportional to $\eta$ corresponds to Ohmic heating, and the term proportional to $\eta_{\rm A}$ corresponds to ambipolar heating. In the single-fluid approximation, the ambipolar heating term replaces the frictional heating term that naturally appears in the multi-fluid formalism (see Equations~(40)–(42) of \citealt{soler2019}).

\subsection{Numerical method}

We adapted the numerical method described in Section~4 of \citet{soler2019} to solve our main Equation~(\ref{eq:main}) using a finite-element-based code. The computational domain roughly corresponds to that shown in Figure~\ref{fig:tube}, although the code only solves in the radial and vertical directions due to the azimuthal invariance of torsional waves. A nonuniform mesh was employed to accurately capture the relevant spatial scales. Convergence tests have been carried out to determine the physically adequate spatial resolution. In the present context, the spatial resolution needs to correctly resolve the wavelength of the Alfv\'enic perturbations along the magnetic field direction and the small transverse scales developed by phase mixing.   A very fine resolution of approximately 100~m is used in the photosphere and lower chromosphere, while the resolution decreases with height, reaching roughly 10~km in the low corona. This resolution is sufficient even for the highest wave frequencies considered here.

A driver for the torsional waves is assumed at the photosphere. Its effect is incorporated through the boundary condition at the lower boundary, where we impose a twist of the magnetic field lines that excites torsional waves with frequencies between 0.1~mHz and 300~mHz. The adopted frequency range is motivated by observations of horizontal motions in the photosphere \citep[see][]{matsushiba2010,matsukitai2010,chitta2012}. Figure~2 of \citet{matsushiba2010} shows that the power spectrum of photospheric horizontal velocity spans approximately from 0.16~mHz to 80~mHz, while the spectrum displayed in Figure~3 of \citet{matsukitai2010} roughly ranges between 0.2~mHz and 20~mHz. In turn, Figure~9 of \citet{chitta2012} covers a range from 0.1~mHz to 100~mHz. Here, the spectrum has been extended up to 300~mHz in order to account for higher frequencies that may not have been detected due to observational limitations. The spectral weighting function is inspired by those used in \citet{tu2013} and \citet{arber2016}, and follows the same power-law form as described in Equation~(66) of \citet{soler2019}. The spectrum is discretized into 84 logarithmically spaced frequencies, with a random phase added to each discrete frequency to construct the full spectrum. The driver amplitude is adjusted so that the horizontally-averaged injected wave energy flux is $10^7$~erg~cm$^{-2}$~s$^{-1}$.  

In contrast to the lower boundary, the upper boundary is treated as perfectly transparent (i.e., non-reflective), allowing all waves that reach it to freely leave the domain. The boundary condition at the lateral edge of the domain, located at $r = r_{\rm max} = 1,000$~km, is that $B'_\varphi$ vanishes.

\begin{figure}
   \centering
\includegraphics[width=0.99\columnwidth]{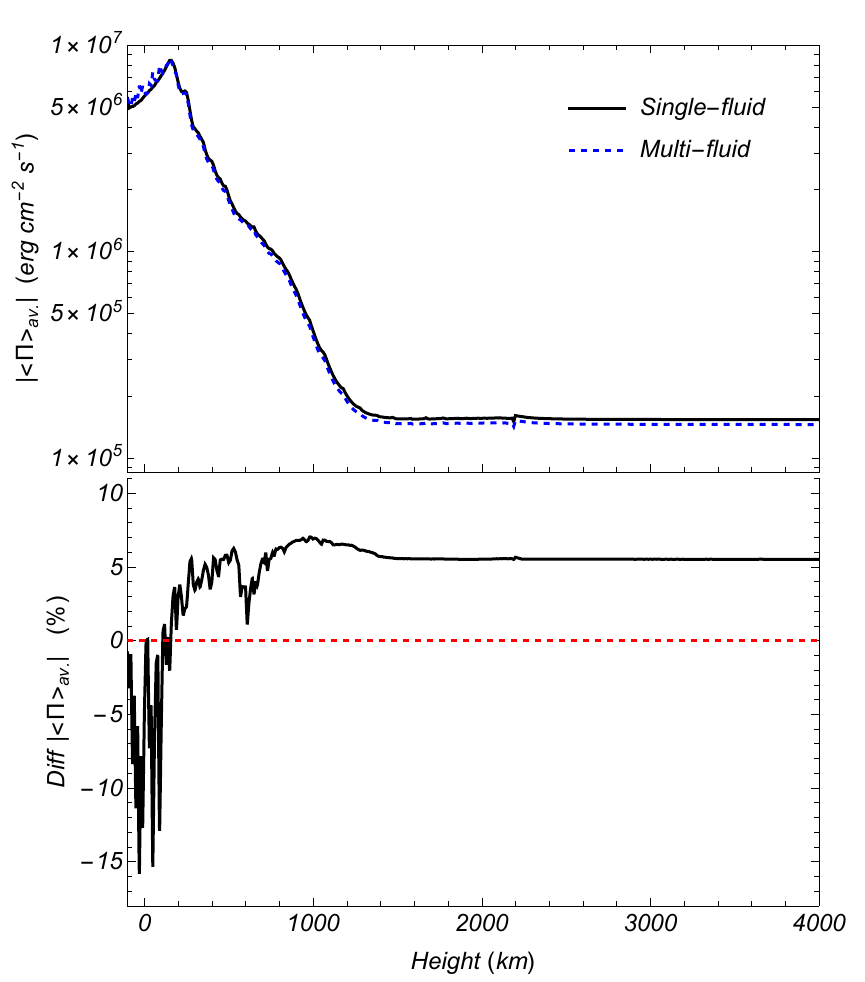} 
      \caption{Dependence on height above the photosphere of the horizontally averaged net energy flux in the single-fluid and multi-fluid models (top) and percentage difference of the single-fluid flux with respect to the multi-fluid flux (bottom).}
         \label{fig:flux1}
   \end{figure}

\section{Comparing single-fluid and multi-fluid results}
\label{sec:res}

We compare the results of the single-fluid model with those of the multi-fluid model. The multi-fluid results are taken from \citet{soler2019} without modification. In contrast, the single-fluid results are new and were specifically computed for this paper. The main Equation~(\ref{eq:main}) is solved for each individual frequency in the spectrum. Then, the time-averaged terms in the wave energy equation (Equations~(\ref{eq:energydensity})--(\ref{eq:heat})) are computed and, subsequently, the results for all the frequencies are added together according to the prescribed spectral weighting function.

First, we are interested in comparing the propagation of wave energy along the vertical direction, $z$. Therefore, to eliminate the dependence on the other two coordinates, $r$ and $\varphi$, we horizontally average the energy flux over an area extending from $r=0$ to $r=r_{\rm max}$ as
\begin{equation}
\left< {\bf \Pi} \right>_{\rm av.} = \frac{1}{\pi r_{\rm max}^2}\int_0^{2\pi} \int_0^{r_{\rm max}} \left< {\bf \Pi} \right>\, r\, {\rm d}r\, {\rm d}\varphi.
\label{eq:average}
\end{equation}
The horizontally averaged flux is a function of height alone. The top panel of Figure~\ref{fig:flux1} displays the horizontally averaged fluxes obtained in the two models. At the scale of the figure, the two results are nearly identical. The net energy flux decreases by several orders of magnitude with height, with only about 1\% of the photospherically injected flux reaching the corona. As shown in \citet{soler2019}, this strong decrease is caused by two mechanisms: reflection and dissipation. Low-frequency waves are mainly reflected back toward the photosphere, with most of the reflection occurring in the middle and upper chromosphere, between about 1{,}000~km above the photosphere and the transition region. High-frequency waves, instead, are efficiently dissipated. Dissipation operates in the lower chromosphere primarily via Ohmic diffusion, and in the middle chromosphere via ion--neutral collisions (or ambipolar diffusion in the single-fluid formalism).

An additional feature is a slight increase in the net flux in the very low chromosphere, up to $\sim200$~km above the photosphere, before the overall decrease sets in. This seemingly counterintuitive behavior arises from the horizontal averaging. While the average injected flux at the photosphere is $10^{7}\,\mathrm{erg\,cm^{-2}\,s^{-1}}$, the small photospheric radius of the tube implies a much larger flux within the tube ($\sim10^{9}\,\mathrm{erg\,cm^{-2}\,s^{-1}}$). Over the first 200~km, reflection and dissipation are still weak, so the upward flux inside the tube remains nearly constant, while the tube radius increases with height due to the expansion. As a result, the horizontally averaged flux increases transiently.

To highlight the differences between the two models, the bottom panel of Figure~\ref{fig:flux1} shows the percentage difference between the single-fluid and multi-fluid energy fluxes. In a narrow region near the photosphere, the single-fluid flux is up to 15\% lower than the multi-fluid flux. At greater heights, this result reverses, and the single-fluid flux becomes slightly larger. At coronal heights, it exceeds the multi-fluid flux by approximately 5\%, which implies that slightly more wave energy reaches the corona in the single-fluid model. 

\begin{figure}
   \centering
\includegraphics[width=0.95\columnwidth]{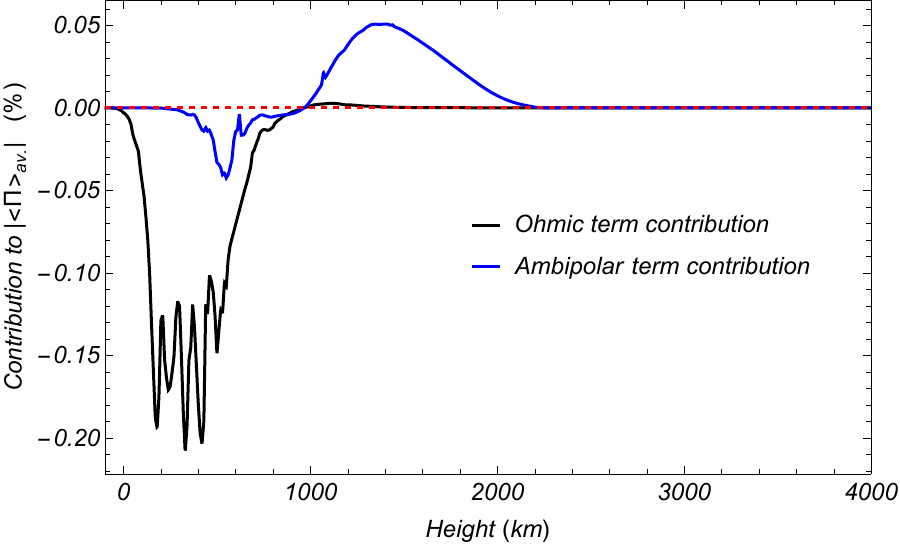} 
      \caption{Percentage contributions of the Ohmic term and the ambipolar term to the horizontally averaged net energy flux as functions of  height above the photosphere. The horizontal red dashed lines denotes no contribution. Results in the single-fluid model.}
         \label{fig:flux2}
   \end{figure}

\begin{figure*}
   \centering
\includegraphics[width=1.65\columnwidth]{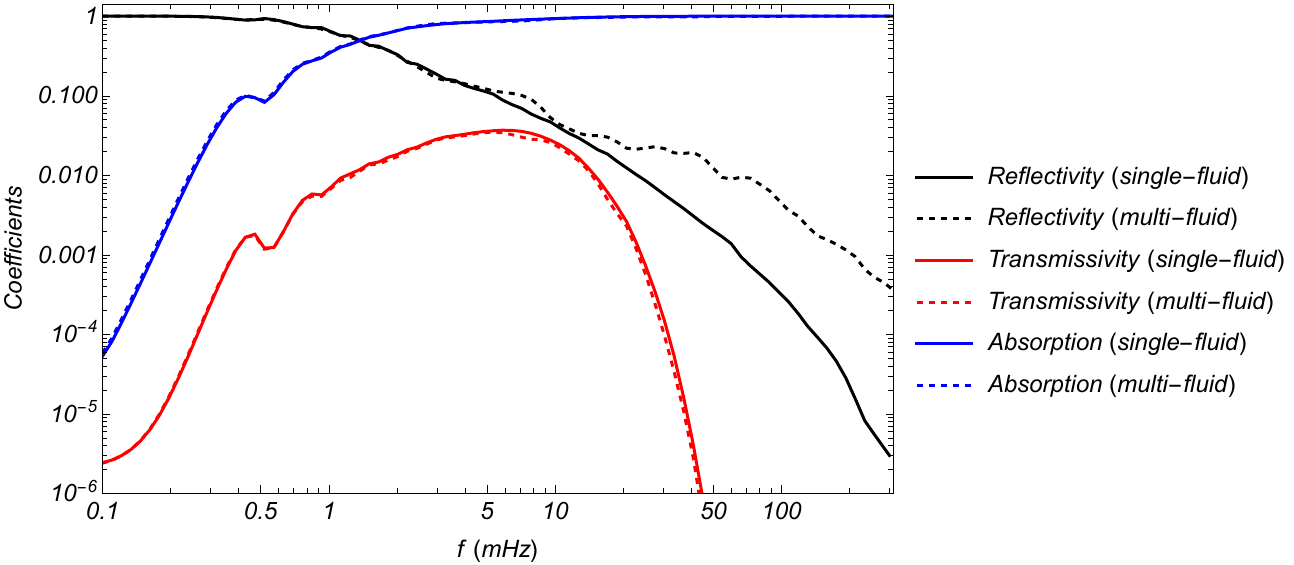} 
      \caption{Coefficients of reflectivity, transmissivity, and absorption as functions of the wave frequency in the single-fluid and multi-fluid models. Note that both axes are in log-scale.}
         \label{fig:coefs}
   \end{figure*}

The nonideal terms were neglected in the computation of the energy flux in the multi-fluid model presented in \citet{soler2019}, but are included in the single-fluid computation. This raises the question of whether the differences between the single-fluid and multi-fluid energy fluxes could be attributed to these terms. To investigate this possibility, Figure~\ref{fig:flux2} displays the percentage contributions of the ambipolar and Ohmic terms relative to the ideal energy flux in the single-fluid model. These nonideal contributions correspond, respectively, to the second and third terms on the right-hand side of Equation~(\ref{eq:energyflux}). The contributions of both terms turn out to be negligible, as assumed by \citet{soler2019}. Specifically, the ambipolar term has a negative contribution of about 0.2\% in the low chromosphere, which seems insufficient to explain the discrepancy between the single-fluid and multi-fluid fluxes at low heights. The Ohmic term has a positive contribution of about 0.05\% in the upper chromosphere, which is again negligible. Importantly, neither term contributes to the energy flux at coronal heights, which clearly indicates that the 5\% excess of energy in the single-fluid model is not related to these terms.

To understand why slightly more energy reaches the corona in the single-fluid model, it is instructive to compute the reflection, transmission, and absorption coefficients following the method explained in Section~3.4 of \citet{soler2019}. The time-averaged energy flux is decomposed into its parallel (upward) and antiparallel (downward) components as $\left< {\bf \Pi} \right> = \left< {\bf \Pi} \right>^\uparrow - \left< {\bf \Pi} \right>^\downarrow$, with
\begin{eqnarray}
\left< {\bf \Pi} \right>^\uparrow &=& \frac{1}{8}\sqrt{\frac{\rho}{\mu}}Z^\uparrow Z^{\uparrow *}\, {\bf B}, \\
\left< {\bf \Pi} \right>^\downarrow &=& \frac{1}{8}\sqrt{\frac{\rho}{\mu}}Z^\downarrow Z^{\downarrow *}\, {\bf B},
\end{eqnarray}
where $\left< {\bf \Pi} \right>^\uparrow$ and $\left< {\bf \Pi} \right>^\downarrow$ correspond to the time-averaged energy fluxes associated to the upward-propagating and downward-propagating Alfv\'en waves, respectively, with $Z^\uparrow$ and $Z^\downarrow$ the Els\"asser variables defined as,
\begin{eqnarray}
Z^\uparrow &=& v'_{\varphi} - \frac{1}{\sqrt{\mu \rho}} B'_\varphi, \\
Z^\downarrow &=& v'_{\varphi} + \frac{1}{\sqrt{\mu \rho}} B'_\varphi.
\end{eqnarray}
We average $\left< {\bf \Pi} \right>^\uparrow$ and $\left< {\bf \Pi} \right>^\downarrow$ horizontally using Equation~(\ref{eq:average}). Then, the relative values of the vertical components of $\left< {\bf \Pi} \right>^\uparrow_{\rm av.}$ and $\left< {\bf \Pi} \right>^\downarrow_{\rm av.}$ at the photospheric and coronal boundaries indicate the fractions of the injected wave energy that are reflected back to the photosphere, transmitted to the corona, or dissipated in between. Therefore, we define the reflection, $\mathcal{R}$, transmission, $\mathcal{T}$, and absorption, $\mathcal{A}$, coefficients as,
\begin{eqnarray}
    \mathcal{R} &=& \frac{\left. \left< {\bf \Pi} \right>^\downarrow_{\rm av.}  \right|_{\rm photosphere} \cdot \hat{e}_z}{\left. \left< {\bf \Pi} \right>^\uparrow_{\rm av.}  \right|_{\rm photosphere} \cdot \hat{e}_z}, \\
    \mathcal{T} &=& \frac{\left. \left< {\bf \Pi} \right>^\uparrow_{\rm av.}  \right|_{\rm corona} \cdot \hat{e}_z}{\left. \left< {\bf \Pi} \right>^\uparrow_{\rm av.}  \right|_{\rm photosphere} \cdot \hat{e}_z}, \\
    \mathcal{A} &=& 1 - \mathcal{R} - \mathcal{T},
\end{eqnarray}
where the absorption coefficient is simply computed by  invoking conservation of energy.

\begin{figure}
   \centering
\includegraphics[width=0.95\columnwidth]{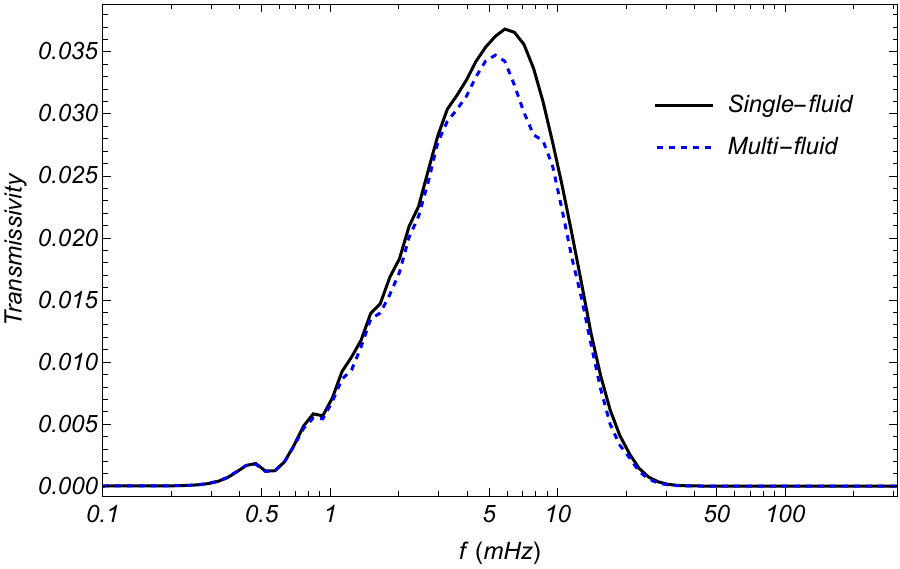} 
      \caption{Transmission coefficient as a function of the wave frequency in the single-fluid and multi-fluid models. Note that the frequency axis is in log-scale.}
         \label{fig:coefs2}
   \end{figure}

Figure~\ref{fig:coefs} displays, on a logarithmic scale, the three coefficients as functions of the driven wave frequency for the two models. The absorption is barely distinguishable between the models and is therefore not discussed further. The transmissivity curves in the two models are also almost identical, but a small difference is appreciable at frequencies around 5~mHz. To better highlight this small discrepancy, Figure~\ref{fig:coefs2} shows both transmissivity curves again, but using a linear scale on the vertical axis. As discussed in \citet{soler2019} and \citet{mortonsoler2025}, the shape of the transmission curve can be approximated by a skewed log-normal function, with a steep tail toward high frequencies and a gentler tail toward low frequencies. This shape is well reproduced by both models, although a discrepancy between them is noticeable around the maximum of the curve, where the single-fluid model yields a slightly larger transmissivity.

Going back to Figure~\ref{fig:coefs}, the most relevant difference between the models appears in the reflectivity at frequencies exceeding approximately 10~mHz. Above this frequency, the reflectivity in the single-fluid model decreases more rapidly with frequency than in the multi-fluid model. The reflection of Alfv\'en waves depends on the relationship between the wavelength and the scale height associated with the gradients of density and magnetic field strength \citep[see, e.g.,][]{ferraro1954}. In the single-fluid model, the relevant density is the total (ion plus neutral) density, since all species are assumed to be strongly coupled regardless of the wave frequency. Conversely, in the multi-fluid model, the effective density felt by the Alfv\'en waves depends on the frequency (see Equation~(34) of \citealt{soler2019}), because the ion--neutral coupling weakens as the frequency increases. Therefore, the density gradient experienced by the Alfv\'en waves becomes progressively different in the single-fluid and multi-fluid models as the frequency increases, becoming slightly steeper in the multi-fluid model. This results in a lower reflectivity in the single-fluid model. In summary, the combined effect of the slightly larger transmissivity around the 5~mHz peak and, more importantly, the lower reflectivity at frequencies higher than 10~mHz can  explain why the energy transmitted to the corona is slightly larger in the single-fluid model.

\begin{figure}
   \centering
\includegraphics[width=0.99\columnwidth]{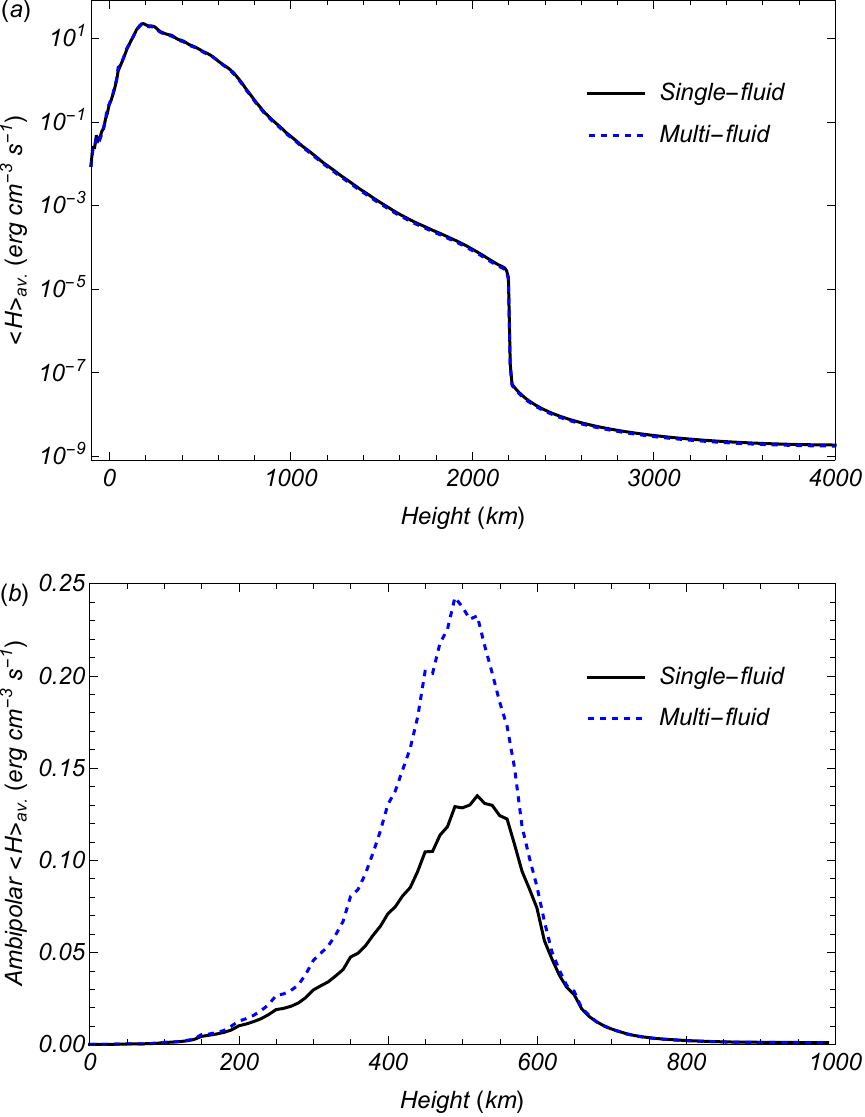} 
      \caption{Dependence on height above the photosphere of the horizontally averaged total heating rate  (top) and ambipolar heating rate (bottom) in the single-fluid and multi-fluid cases. Note that the bottom panel only displays heights below 1,000~km.}
         \label{fig:heating}
   \end{figure}

Finally, we examine the differences in the heating rate associated with the dissipation of wave energy. As in the case of the energy flux, we consider the horizontally averaged heating rate computed using an expression equivalent to Equation~(\ref{eq:average}). The top panel of Figure~\ref{fig:heating} shows the averaged total heating rate as a function of height above the photosphere. Once again, it is remarkable how similar the results are for the two models, which appear nearly superimposed in the figure. According to Equation~(\ref{eq:heat}), the total heating rate is the sum of the Ohmic and ambipolar contributions, with frictional heating replacing ambipolar heating in the multi-fluid description (see Equation~(58) of \citealt{soler2019}). We have compared these contributions separately for both models. The Ohmic heating is, once more, nearly identical in the two models, and its comparison is not shown here for the sake of simplicity. The ambipolar and frictional heating rates are also extremely similar at all heights, with the exception of a narrow region around 500~km above the photosphere. The bottom panel of Figure~\ref{fig:heating} compares the ambipolar and frictional heating rates of the two models in the vicinity of this region, where it can be seen that the single-fluid model underestimates the heating rate by approximately a factor of two compared with the multi-fluid result. The reason for this discrepancy lies in the approximate description of the ion-neutral drifts in the single-fluid model. For the multi-fluid model, Figure~7a of \citet{soler2019} shows the averaged ion-neutral drift as a function of height at three different radial positions in the flux tube. The height at which these drifts attain their largest values in the chromosphere coincides with the region of discrepancy discussed here. This height also corresponds to the location in the atmosphere where the neutral-ion collision frequencies reach their minimum (see again Figure~\ref{fig:model}). This clearly indicates that the drifts are only partially captured by the single-fluid approximation, which leads to an underestimation of the associated heating rate in the single-fluid model.

\section{Discussion}
\label{sec:conc}

The purpose of this paper has been to compare the results of multi-fluid and single-fluid models for the propagation of torsional Alfv\'en waves from the photosphere to the corona along a vertical magnetic flux tube that expands with height. The results in the multi-fluid model were previously obtained in \citet{soler2019}. Overall, the differences between the two models are negligible, with only two noticeable discrepancies. First, the single-fluid model allows slightly more wave energy (about 5\% more) to reach the corona than the multi-fluid model. This difference is explained by the lower reflectivity at high frequencies and the slightly higher transmissivity around the 5~mHz peak obtained in the single-fluid model. Second, in a layer located around 500~km above the photosphere, the ambipolar heating rate predicted by the single-fluid model is approximately a factor of two smaller than the corresponding frictional heating rate in the multi-fluid model. This discrepancy arises from the approximate treatment of ion-neutral drifts in the single-fluid formulation, which become significant at those heights. In summary, for the description of Alfv\'en waves in the lower solar atmosphere, the single-fluid approximation yields remarkably accurate results when compared with the more physically applicable multi-fluid model.

Recently, \citet{gomez2025} compared single-fluid and multi-fluid results for MHD wave propagation in the lower solar atmosphere, so that their study and the present paper share a common goal. However, the background configuration, methodology, and type of wave studied differ. \citet{gomez2025} considered fast magneto-acoustic waves in a stratified atmosphere of pure hydrogen with a homogeneous horizontal magnetic field, whereas we studied torsional Alfv\'en waves in a hydrogen-helium stratified atmosphere with a vertical flux tube expanding with height. They employed weakly nonlinear, time-dependent simulations complemented by analytical approximations based on the eikonal approach, while we numerically solved the linearized equations in the stationary state. Here we considered the effects of Ohm's and ambipolar diffusion, while \citet{gomez2025} only included ambipolar diffusion.  Despite these differences, a particular finding of \citet{gomez2025} can be meaningfully compared with our results. Similar to what we find here, \citet{gomez2025} showed that fast waves in the multi-fluid model transport less energy upward than in the single-fluid model. They explained this behavior by noting that the multi-fluid model is more efficient at dissipating energy at low heights than the single-fluid model. This result qualitatively agrees with Figure~\ref{fig:heating}(b), although in our case it is not the main mechanism responsible for the reduced Alfv\'en wave energy transport to the corona in the multi-fluid model. Instead, we attribute this difference to variations in the atmospheric reflectivity that appear at high frequencies, when the effective density felt by the Alfv\'en waves in the multi-fluid model begins to differ from the total density.

Another difference with the study by \citet{gomez2025} is that the Alfv\'en waves investigated here are incompressible, whereas fast magneto-acoustic waves are compressible. Consequently, all effects associated with compressibility, pressure variations, and related phenomena discussed in \citet{gomez2025} are not applicable to the case of Alfv\'en waves.

Finally, it is worth noting that the main conclusion of this work, namely, the excellent accuracy of the single-fluid model, is not generalizable beyond the specific context of the solar atmospheric Alfv\'en waves studied here. Processes such as shocks and instabilities in partially ionized solar and astrophysical plasmas exhibit distinctive features when investigated with multi-fluid models that cannot be reproduced by the single-fluid approximation \citep[see][and references therein]{solerballester2022,hilliersnow2023}. We also note that the highest wave frequency included in our analysis was 300~mHz, and that more significant differences between the models may emerge if higher frequencies are considered \citep[see, e.g.,][]{soler2013,martinezgomez2025}.  However, observations by \citet{matsushiba2010}, \citet{matsukitai2010}, and \citet{chitta2012} indicate that the power of horizontal photospheric motions decreases rapidly toward higher frequencies. Therefore, it is unlikely that very high-frequency Alfv\'en waves are driven in the photosphere with sufficient energy to play a significant role.

\begin{acknowledgements}
      This publication is part of the R+D+i project PID2023-147708NB-I00, funded by MCIN/AEI/10.13039/501100011033 and by FEDER, EU. The author acknowledges David Mart\'inez-G\'omez for a useful discussion and an anonymous referee for a constructive report.
\end{acknowledgements}

   \bibliographystyle{aa} 
   \bibliography{refs}

@ARTICLE{soler2019,
       author = {{Soler}, Roberto and {Terradas}, Jaume and {Oliver}, Ram{\'o}n and
         {Ballester}, Jos{\'e} Luis},
        title = "{Energy Transport and Heating by Torsional Alfv{\'e}n Waves Propagating from the Photosphere to the Corona in the Quiet Sun}",
      journal = {\apj},
         year = "2019",
        month = "Jan",
       volume = {871},
       number = {1},
          eid = {3},
        pages = {3},
          doi = {10.3847/1538-4357/aaf64c},
archivePrefix = {arXiv},
       eprint = {1812.01323},
 primaryClass = {astro-ph.SR},
       adsurl = {https://ui.adsabs.harvard.edu/abs/2019ApJ...871....3S}
}

@ARTICLE{FAL93,
       author = {{Fontenla}, J.~M. and {Avrett}, E.~H. and {Loeser}, R.},
        title = "{Energy Balance in the Solar Transition Region. III. Helium Emission in Hydrostatic, Constant-Abundance Models with Diffusion}",
      journal = {\apj},
         year = "1993",
        month = "Mar",
       volume = {406},
        pages = {319},
          doi = {10.1086/172443},
       adsurl = {https://ui.adsabs.harvard.edu/abs/1993ApJ...406..319F}
}

@ARTICLE{zaqarashvili2013,
       author = {{Zaqarashvili}, T.~V. and {Khodachenko}, M.~L. and {Soler}, R.},
        title = "{Torsional Alfv{\'e}n waves in partially ionized solar plasma: effects of neutral helium and stratification}",
      journal = {\aap},
         year = "2013",
        month = "Jan",
       volume = {549},
          eid = {A113},
        pages = {A113},
          doi = {10.1051/0004-6361/201220272},
archivePrefix = {arXiv},
       eprint = {1211.1348},
 primaryClass = {astro-ph.SR},
       adsurl = {https://ui.adsabs.harvard.edu/abs/2013A&A...549A.113Z}
}

@BOOK{walker2005,
   author = {{Walker}, A.~D.~M.},
    title = "{Magnetohydrodynamic Waves in Geospace}",
     year = 2005,
  isbn = {9780750309103},
  series = {Series in Plasma Physics,},
  publisher = {Institute of Physics Publishing}    
}

@ARTICLE{khomenko2012,
       author = {{Khomenko}, E. and {Collados}, M.},
        title = "{Heating of the Magnetized Solar Chromosphere by Partial Ionization Effects}",
      journal = {\apj},
         year = "2012",
        month = "Mar",
       volume = {747},
       number = {2},
          eid = {87},
        pages = {87},
          doi = {10.1088/0004-637X/747/2/87},
archivePrefix = {arXiv},
       eprint = {1112.3374},
 primaryClass = {astro-ph.SR},
       adsurl = {https://ui.adsabs.harvard.edu/abs/2012ApJ...747...87K}
}

@ARTICLE{khomenko2014,
       author = {{Khomenko}, E. and {Collados}, M. and {D{\'\i}az}, A. and {Vitas}, N.},
        title = "{Fluid description of multi-component solar partially ionized plasma}",
      journal = {Physics of Plasmas},
         year = "2014",
        month = "Sep",
       volume = {21},
       number = {9},
          eid = {092901},
        pages = {092901},
          doi = {10.1063/1.4894106},
archivePrefix = {arXiv},
       eprint = {1408.1871},
 primaryClass = {astro-ph.SR},
       adsurl = {https://ui.adsabs.harvard.edu/abs/2014PhPl...21i2901K}
}

@ARTICLE{ballester2018,
       author = {{Ballester}, Jos{\'e} Luis and {Alexeev}, Igor and {Collados}, Manuel and
         {Downes}, Turlough and {Pfaff}, Robert F. and {Gilbert}, Holly and
         {Khodachenko}, Maxim and {Khomenko}, Elena and
         {Shaikhislamov}, Ildar F. and {Soler}, Roberto and
         {V{\'a}zquez-Semadeni}, Enrique and {Zaqarashvili}, Teimuraz},
        title = "{Partially Ionized Plasmas in Astrophysics}",
      journal = {\ssr},
         year = "2018",
        month = "Mar",
       volume = {214},
       number = {2},
          eid = {58},
        pages = {58},
          doi = {10.1007/s11214-018-0485-6},
archivePrefix = {arXiv},
       eprint = {1707.07975},
 primaryClass = {astro-ph.SR},
       adsurl = {https://ui.adsabs.harvard.edu/abs/2018SSRv..214...58B}
}

@ARTICLE{arber2016,
       author = {{Arber}, T.~D. and {Brady}, C.~S. and {Shelyag}, S.},
        title = "{Alfv{\'e}n Wave Heating of the Solar Chromosphere: 1.5D Models}",
      journal = {\apj},
         year = 2016,
        month = feb,
       volume = {817},
       number = {2},
          eid = {94},
        pages = {94},
          doi = {10.3847/0004-637X/817/2/94},
archivePrefix = {arXiv},
       eprint = {1512.05816},
 primaryClass = {astro-ph.SR},
       adsurl = {https://ui.adsabs.harvard.edu/abs/2016ApJ...817...94A}
}

@ARTICLE{tu2013,
       author = {{Tu}, Jiannan and {Song}, Paul},
        title = "{A Study of Alfv{\'e}n Wave Propagation and Heating the Chromosphere}",
      journal = {\apj},
         year = 2013,
        month = nov,
       volume = {777},
       number = {1},
          eid = {53},
        pages = {53},
          doi = {10.1088/0004-637X/777/1/53},
       adsurl = {https://ui.adsabs.harvard.edu/abs/2013ApJ...777...53T}
}

@ARTICLE{hillier2024,
       author = {{Hillier}, Andrew S.},
        title = "{On the ambipolar diffusion formulation for ion-neutral drifts in the non-negligible drift velocity limit}",
      journal = {Philosophical Transactions of the Royal Society of London Series A},
         year = 2024,
        month = jun,
       volume = {382},
       number = {2272},
          eid = {20230229},
        pages = {20230229},
          doi = {10.1098/rsta.2023.0229},
archivePrefix = {arXiv},
       eprint = {2403.16847},
 primaryClass = {astro-ph.SR},
       adsurl = {https://ui.adsabs.harvard.edu/abs/2024RSPTA.38230229H}
}

@ARTICLE{soler2025,
       author = {{Soler}, Roberto},
        title = "{Alfv{\'e}n wave propagation from the photosphere to the corona: Temporal evolution against stationary results}",
      journal = {\aap},
         year = 2025,
        month = sep,
       volume = {702},
          eid = {A16},
        pages = {A16},
          doi = {10.1051/0004-6361/202556264},
archivePrefix = {arXiv},
       eprint = {2508.21758},
 primaryClass = {astro-ph.SR},
       adsurl = {https://ui.adsabs.harvard.edu/abs/2025A&A...702A..16S}
}

@ARTICLE{ferraro1954,
       author = {{Ferraro}, V.~C.~A.},
        title = "{On the Reflection and Refraction of Alfven Waves.}",
      journal = {\apj},
         year = 1954,
        month = mar,
       volume = {119},
        pages = {393},
          doi = {10.1086/145837},
       adsurl = {https://ui.adsabs.harvard.edu/abs/1954ApJ...119..393F}
}

@ARTICLE{gomez2025,
       author = {{G{\'o}mez-M{\'\i}guez}, M.~M. and {Mart{\'\i}nez-G{\'o}mez}, D. and {Khomenko}, E. and {Popescu Braileanu}, B. and {Collados}, M. and {Cally}, P.~S.},
        title = "{Fast magneto-acoustic waves in the solar chromosphere: Comparison of single-fluid and two-fluid approximations}",
      journal = {\aap},
         year = 2025,
        month = sep,
       volume = {701},
          eid = {A196},
        pages = {A196},
          doi = {10.1051/0004-6361/202554637},
archivePrefix = {arXiv},
       eprint = {2507.16460},
 primaryClass = {astro-ph.SR},
       adsurl = {https://ui.adsabs.harvard.edu/abs/2025A&A...701A.196G}
}

@ARTICLE{morton2026,
       author = {{Morton}, R.~J. and {Gao}, Y. and {Tajfirouze}, E. and {Tian}, H. and {Van Doorsselaere}, T. and {Schad}, T.~A.},
        title = "{Evidence for small-scale torsional Alfv{\'e}n waves in the solar corona}",
      journal = {Nature Astronomy},
         year = 2026,
        month = jan,
       volume = {10},
        pages = {42-53},
          doi = {10.1038/s41550-025-02690-9},
       adsurl = {https://ui.adsabs.harvard.edu/abs/2026NatAs..10...42M}
}

@ARTICLE{mortonsoler2025,
       author = {{Morton}, Richard J. and {Soler}, Roberto},
        title = "{On the Origins of Coronal Alfv{\'e}nic Waves}",
      journal = {\apjl},
         year = 2025,
        month = jun,
       volume = {986},
       number = {1},
          eid = {L6},
        pages = {L6},
          doi = {10.3847/2041-8213/add7da},
archivePrefix = {arXiv},
       eprint = {2505.08636},
 primaryClass = {astro-ph.SR},
       adsurl = {https://ui.adsabs.harvard.edu/abs/2025ApJ...986L...6M}
}

@ARTICLE{solerballester2022,
       author = {{Soler}, Roberto and {Ballester}, Jos{\'e} Luis},
        title = "{Theory of Fluid Instabilities in Partially Ionized Plasmas: An Overview}",
      journal = {Frontiers in Astronomy and Space Sciences},
         year = 2022,
        month = may,
       volume = {9},
          eid = {789083},
        pages = {789083},
          doi = {10.3389/fspas.2022.789083},
       adsurl = {https://ui.adsabs.harvard.edu/abs/2022FrASS...9.9083S}
}

@ARTICLE{hilliersnow2023,
       author = {{Hillier}, Andrew and {Snow}, Ben},
        title = "{Shocks and instabilities in the partially ionised solar atmosphere}",
      journal = {Advances in Space Research},
         year = 2023,
        month = feb,
       volume = {71},
       number = {4},
        pages = {1962-1983},
          doi = {10.1016/j.asr.2022.08.079},
archivePrefix = {arXiv},
       eprint = {2302.07362},
 primaryClass = {astro-ph.SR},
       adsurl = {https://ui.adsabs.harvard.edu/abs/2023AdSpR..71.1962H}
}

@ARTICLE{martinezgomez2025,
       author = {{Mart{\'\i}nez-G{\'o}mez}, David},
        title = "{Propagation of Waves in Weakly Ionized Two-fluid Plasmas. I. Small-amplitude Alfv{\'e}nic Waves}",
      journal = {\apj},
         year = 2025,
        month = mar,
       volume = {982},
       number = {1},
          eid = {4},
        pages = {4},
          doi = {10.3847/1538-4357/adb713},
archivePrefix = {arXiv},
       eprint = {2502.13659},
 primaryClass = {astro-ph.SR},
       adsurl = {https://ui.adsabs.harvard.edu/abs/2025ApJ...982....4M}
}

@ARTICLE{soler2013,
       author = {{Soler}, R. and {Carbonell}, M. and {Ballester}, J.~L. and {Terradas}, J.},
        title = "{Alfv{\'e}n Waves in a Partially Ionized Two-fluid Plasma}",
      journal = {\apj},
         year = 2013,
        month = apr,
       volume = {767},
       number = {2},
          eid = {171},
        pages = {171},
          doi = {10.1088/0004-637X/767/2/171},
archivePrefix = {arXiv},
       eprint = {1303.4297},
 primaryClass = {astro-ph.SR},
       adsurl = {https://ui.adsabs.harvard.edu/abs/2013ApJ...767..171S}
}

@ARTICLE{soler2024,
       author = {{Soler}, Roberto},
        title = "{Magnetohydrodynamic waves in the partially ionized solar plasma}",
      journal = {Philosophical Transactions of the Royal Society of London Series A},
         year = 2024,
        month = jun,
       volume = {382},
       number = {2272},
          eid = {20230223},
        pages = {20230223},
          doi = {10.1098/rsta.2023.0223},
       adsurl = {https://ui.adsabs.harvard.edu/abs/2024RSPTA.38230223S}
}

@ARTICLE{zaqarashvili2011,
       author = {{Zaqarashvili}, T.~V. and {Khodachenko}, M.~L. and {Rucker}, H.~O.},
        title = "{Magnetohydrodynamic waves in solar partially ionized plasmas: two-fluid approach}",
      journal = {\aap},
         year = 2011,
        month = may,
       volume = {529},
          eid = {A82},
        pages = {A82},
          doi = {10.1051/0004-6361/201016326},
archivePrefix = {arXiv},
       eprint = {1101.3913},
 primaryClass = {astro-ph.SR},
       adsurl = {https://ui.adsabs.harvard.edu/abs/2011A&A...529A..82Z}
}

@ARTICLE{diaz2021,
       author = {{D{\'\i}az-Su{\'a}rez}, Sergio and {Soler}, Roberto},
        title = "{Transition to turbulence in nonuniform coronal loops driven by torsional Alfv{\'e}n waves}",
      journal = {\aap},
         year = 2021,
        month = apr,
       volume = {648},
          eid = {A22},
        pages = {A22},
          doi = {10.1051/0004-6361/202040161},
archivePrefix = {arXiv},
       eprint = {2102.06464},
 primaryClass = {astro-ph.SR},
       adsurl = {https://ui.adsabs.harvard.edu/abs/2021A&A...648A..22D}
}

@ARTICLE{guo2019,
       author = {{Guo}, Mingzhe and {Van Doorsselaere}, Tom and {Karampelas}, Konstantinos and {Li}, Bo and {Antolin}, Patrick and {De Moortel}, Ineke},
        title = "{Heating Effects from Driven Transverse and Alfv{\'e}n Waves in Coronal Loops}",
      journal = {\apj},
         year = 2019,
        month = jan,
       volume = {870},
       number = {2},
          eid = {55},
        pages = {55},
          doi = {10.3847/1538-4357/aaf1d0},
archivePrefix = {arXiv},
       eprint = {1811.07608},
 primaryClass = {astro-ph.SR},
       adsurl = {https://ui.adsabs.harvard.edu/abs/2019ApJ...870...55G}
}

@ARTICLE{depontieu2014,
       author = {{De Pontieu}, B. and {Rouppe van der Voort}, L. and {McIntosh}, S.~W. and {Pereira}, T.~M.~D. and {Carlsson}, M. and {Hansteen}, V. and {Skogsrud}, H. and {Lemen}, J. and {Title}, A. and {Boerner}, P. and {Hurlburt}, N. and {Tarbell}, T.~D. and {Wuelser}, J.~P. and {De Luca}, E.~E. and {Golub}, L. and {McKillop}, S. and {Reeves}, K. and {Saar}, S. and {Testa}, P. and {Tian}, H. and {Kankelborg}, C. and {Jaeggli}, S. and {Kleint}, L. and {Martinez-Sykora}, J.},
        title = "{On the prevalence of small-scale twist in the solar chromosphere and transition region}",
      journal = {Science},
         year = 2014,
        month = oct,
       volume = {346},
       number = {6207},
          eid = {1255732},
        pages = {1255732},
          doi = {10.1126/science.1255732},
archivePrefix = {arXiv},
       eprint = {1410.6862},
 primaryClass = {astro-ph.SR},
       adsurl = {https://ui.adsabs.harvard.edu/abs/2014Sci...346D.315D}
}

@ARTICLE{depontieu2012,
       author = {{De Pontieu}, B. and {Carlsson}, M. and {Rouppe van der Voort}, L.~H.~M. and {Rutten}, R.~J. and {Hansteen}, V.~H. and {Watanabe}, H.},
        title = "{Ubiquitous Torsional Motions in Type II Spicules}",
      journal = {\apjl},
         year = 2012,
        month = jun,
       volume = {752},
       number = {1},
          eid = {L12},
        pages = {L12},
          doi = {10.1088/2041-8205/752/1/L12},
archivePrefix = {arXiv},
       eprint = {1205.5006},
 primaryClass = {astro-ph.SR},
       adsurl = {https://ui.adsabs.harvard.edu/abs/2012ApJ...752L..12D}
}

@ARTICLE{jess2009,
       author = {{Jess}, David B. and {Mathioudakis}, Mihalis and {Erd{\'e}lyi}, Robert and {Crockett}, Philip J. and {Keenan}, Francis P. and {Christian}, Damian J.},
        title = "{Alfv{\'e}n Waves in the Lower Solar Atmosphere}",
      journal = {Science},
         year = 2009,
        month = mar,
       volume = {323},
       number = {5921},
        pages = {1582},
          doi = {10.1126/science.1168680},
archivePrefix = {arXiv},
       eprint = {0903.3546},
 primaryClass = {astro-ph.SR},
       adsurl = {https://ui.adsabs.harvard.edu/abs/2009Sci...323.1582J}
}

@ARTICLE{alfven1947,
       author = {{Alfv{\'e}n}, H.},
        title = "{Magneto hydrodynamic waves, and the heating of the solar corona}",
      journal = {\mnras},
         year = 1947,
        month = jan,
       volume = {107},
        pages = {211},
          doi = {10.1093/mnras/107.2.211},
       adsurl = {https://ui.adsabs.harvard.edu/abs/1947MNRAS.107..211A}
}

@ARTICLE{erdelyi2007,
       author = {{Erd{\'e}lyi}, R. and {Fedun}, V.},
        title = "{Are There Alfv{\'e}n Waves in the Solar Atmosphere?}",
      journal = {Science},
         year = 2007,
        month = dec,
       volume = {318},
       number = {5856},
        pages = {1572},
          doi = {10.1126/science.1153006},
       adsurl = {https://ui.adsabs.harvard.edu/abs/2007Sci...318.1572E}
}

@ARTICLE{soler2017,
       author = {{Soler}, Roberto and {Terradas}, Jaume and {Oliver}, Ram{\'o}n and {Ballester}, Jos{\'e} Luis},
        title = "{Propagation of Torsional Alfv{\'e}n Waves from the Photosphere to the Corona: Reflection, Transmission, and Heating in Expanding Flux Tubes}",
      journal = {\apj},
         year = 2017,
        month = may,
       volume = {840},
       number = {1},
          eid = {20},
        pages = {20},
          doi = {10.3847/1538-4357/aa6d7f},
       adsurl = {https://ui.adsabs.harvard.edu/abs/2017ApJ...840...20S}
}

@ARTICLE{forteza2007,
       author = {{Forteza}, P. and {Oliver}, R. and {Ballester}, J.~L. and {Khodachenko}, M.~L.},
        title = "{Damping of oscillations by ion-neutral collisions in a prominence plasma}",
      journal = {\aap},
         year = 2007,
        month = jan,
       volume = {461},
       number = {2},
        pages = {731-739},
          doi = {10.1051/0004-6361:20065900},
       adsurl = {https://ui.adsabs.harvard.edu/abs/2007A&A...461..731F}
}

@ARTICLE{goodman2011,
       author = {{Goodman}, Michael L.},
        title = "{Conditions for Photospherically Driven Alfv{\'e}nic Oscillations to Heat the Solar Chromosphere by Pedersen Current Dissipation}",
      journal = {\apj},
         year = 2011,
        month = jul,
       volume = {735},
       number = {1},
          eid = {45},
        pages = {45},
          doi = {10.1088/0004-637X/735/1/45},
archivePrefix = {arXiv},
       eprint = {1410.8519},
 primaryClass = {astro-ph.SR},
       adsurl = {https://ui.adsabs.harvard.edu/abs/2011ApJ...735...45G}
}

@ARTICLE{sykora2015,
       author = {{Mart{\'\i}nez-Sykora}, Juan and {De Pontieu}, Bart and {Hansteen}, Viggo and {Carlsson}, Mats},
        title = "{The role of partial ionization effects in the chromosphere}",
      journal = {Philosophical Transactions of the Royal Society of London Series A},
         year = 2015,
        month = apr,
       volume = {373},
       number = {2042},
        pages = {20140268-20140268},
          doi = {10.1098/rsta.2014.0268},
archivePrefix = {arXiv},
       eprint = {1503.02723},
 primaryClass = {astro-ph.SR},
       adsurl = {https://ui.adsabs.harvard.edu/abs/2015RSPTA.37340268M}
}

@ARTICLE{Khodachenko2004,
       author = {{Khodachenko}, M.~L. and {Arber}, T.~D. and {Rucker}, H.~O. and {Hanslmeier}, A.},
        title = "{Collisional and viscous damping of MHD waves in partially ionized plasmas of the solar atmosphere}",
      journal = {\aap},
         year = 2004,
        month = aug,
       volume = {422},
        pages = {1073-1084},
          doi = {10.1051/0004-6361:20034207},
       adsurl = {https://ui.adsabs.harvard.edu/abs/2004A&A...422.1073K}
}

@ARTICLE{Leake2005,
       author = {{Leake}, J.~E. and {Arber}, T.~D. and {Khodachenko}, M.~L.},
        title = "{Collisional dissipation of Alfv{\'e}n waves in a partially ionised solar chromosphere}",
      journal = {\aap},
         year = 2005,
        month = nov,
       volume = {442},
       number = {3},
        pages = {1091-1098},
          doi = {10.1051/0004-6361:20053427},
archivePrefix = {arXiv},
       eprint = {astro-ph/0510265},
 primaryClass = {astro-ph},
       adsurl = {https://ui.adsabs.harvard.edu/abs/2005A&A...442.1091L}
}

@ARTICLE{cally2019,
       author = {{Cally}, Paul S. and {Khomenko}, Elena},
        title = "{Fast-to-Alfv{\'e}n Mode Conversion and Ambipolar Heating in Structured Media. I. Simplified Cold Plasma Model}",
      journal = {\apj},
         year = 2019,
        month = nov,
       volume = {885},
       number = {1},
          eid = {58},
        pages = {58},
          doi = {10.3847/1538-4357/ab3bce},
       adsurl = {https://ui.adsabs.harvard.edu/abs/2019ApJ...885...58C}
}

@ARTICLE{song2011,
       author = {{Song}, P. and {Vasyli{\={u}}nas}, V.~M.},
        title = "{Heating of the solar atmosphere by strong damping of Alfv{\'e}n waves}",
      journal = {Journal of Geophysical Research (Space Physics)},
         year = 2011,
        month = sep,
       volume = {116},
       number = {A9},
          eid = {A09104},
        pages = {A09104},
          doi = {10.1029/2011JA016679},
       adsurl = {https://ui.adsabs.harvard.edu/abs/2011JGRA..116.9104S}
}

@ARTICLE{popescu2019,
       author = {{Popescu Braileanu}, B. and {Lukin}, V.~S. and {Khomenko}, E. and {de Vicente}, {\'A}.},
        title = "{Two-fluid simulations of waves in the solar chromosphere. I. Numerical code verification}",
      journal = {\aap},
         year = 2019,
        month = jul,
       volume = {627},
          eid = {A25},
        pages = {A25},
          doi = {10.1051/0004-6361/201834154},
archivePrefix = {arXiv},
       eprint = {1905.03559},
 primaryClass = {astro-ph.SR},
       adsurl = {https://ui.adsabs.harvard.edu/abs/2019A&A...627A..25P}
}

@ARTICLE{Pelekhata2021,
       author = {{Pelekhata}, M. and {Murawski}, K. and {Poedts}, S.},
        title = "{Solar chromosphere heating and generation of plasma outflows by impulsively generated two-fluid Alfv{\'e}n waves}",
      journal = {\aap},
         year = 2021,
        month = aug,
       volume = {652},
          eid = {A114},
        pages = {A114},
          doi = {10.1051/0004-6361/202141262},
archivePrefix = {arXiv},
       eprint = {2107.12032},
 primaryClass = {astro-ph.SR},
       adsurl = {https://ui.adsabs.harvard.edu/abs/2021A&A...652A.114P}
}

@ARTICLE{martinez2018,
       author = {{Mart{\'\i}nez-G{\'o}mez}, David and {Soler}, Roberto and {Terradas}, Jaume},
        title = "{Multi-fluid Approach to High-frequency Waves in Plasmas. III. Nonlinear Regime and Plasma Heating}",
      journal = {\apj},
         year = 2018,
        month = mar,
       volume = {856},
       number = {1},
          eid = {16},
        pages = {16},
          doi = {10.3847/1538-4357/aab156},
archivePrefix = {arXiv},
       eprint = {1802.08134},
 primaryClass = {astro-ph.SR},
       adsurl = {https://ui.adsabs.harvard.edu/abs/2018ApJ...856...16M}
}

@ARTICLE{martinez2017,
       author = {{Mart{\'\i}nez-G{\'o}mez}, David and {Soler}, Roberto and {Terradas}, Jaume},
        title = "{Multi-fluid Approach to High-frequency Waves in Plasmas. II. Small-amplitude Regime in Partially Ionized Media}",
      journal = {\apj},
         year = 2017,
        month = mar,
       volume = {837},
       number = {1},
          eid = {80},
        pages = {80},
          doi = {10.3847/1538-4357/aa5eab},
archivePrefix = {arXiv},
       eprint = {1703.05093},
 primaryClass = {astro-ph.SR},
       adsurl = {https://ui.adsabs.harvard.edu/abs/2017ApJ...837...80M}
}

@ARTICLE{cally2023,
       author = {{Cally}, Paul S.},
        title = "{Efficiency of Magnetohydrodynamic Wave Generation in Weakly Ionized Atmospheres}",
      journal = {\apj},
         year = 2023,
        month = sep,
       volume = {954},
       number = {1},
          eid = {85},
        pages = {85},
          doi = {10.3847/1538-4357/acdc99},
archivePrefix = {arXiv},
       eprint = {2306.04801},
 primaryClass = {astro-ph.SR},
       adsurl = {https://ui.adsabs.harvard.edu/abs/2023ApJ...954...85C}
}

@ARTICLE{maneva2017,
       author = {{Maneva}, Yana G. and {Alvarez Laguna}, Alejandro and {Lani}, Andrea and {Poedts}, Stefaan},
        title = "{Multi-fluid Modeling of Magnetosonic Wave Propagation in the Solar Chromosphere: Effects of Impact Ionization and Radiative Recombination}",
      journal = {\apj},
         year = 2017,
        month = feb,
       volume = {836},
       number = {2},
          eid = {197},
        pages = {197},
          doi = {10.3847/1538-4357/aa5b83},
archivePrefix = {arXiv},
       eprint = {1611.08439},
 primaryClass = {astro-ph.SR},
       adsurl = {https://ui.adsabs.harvard.edu/abs/2017ApJ...836..197M}
}

@ARTICLE{chitta2012,
       author = {{Chitta}, L.~P. and {van Ballegooijen}, A.~A. and {Rouppe van der Voort}, L. and {DeLuca}, E.~E. and {Kariyappa}, R.},
        title = "{Dynamics of the Solar Magnetic Bright Points Derived from Their Horizontal Motions}",
      journal = {\apj},
         year = 2012,
        month = jun,
       volume = {752},
       number = {1},
          eid = {48},
        pages = {48},
          doi = {10.1088/0004-637X/752/1/48},
archivePrefix = {arXiv},
       eprint = {1204.4362},
 primaryClass = {astro-ph.SR},
       adsurl = {https://ui.adsabs.harvard.edu/abs/2012ApJ...752...48C}
}

@ARTICLE{matsushiba2010,
       author = {{Matsumoto}, Takuma and {Shibata}, Kazunari},
        title = "{Nonlinear Propagation of Alfv{\'e}n Waves Driven by Observed Photospheric Motions: Application to the Coronal Heating and Spicule Formation}",
      journal = {\apj},
         year = 2010,
        month = feb,
       volume = {710},
       number = {2},
        pages = {1857-1867},
          doi = {10.1088/0004-637X/710/2/1857},
archivePrefix = {arXiv},
       eprint = {1001.4307},
 primaryClass = {astro-ph.SR},
       adsurl = {https://ui.adsabs.harvard.edu/abs/2010ApJ...710.1857M}
}

@ARTICLE{matsukitai2010,
       author = {{Matsumoto}, Takuma and {Kitai}, Reizaburo},
        title = "{Temporal Power Spectra of the Horizontal Velocity of the Solar Photosphere}",
      journal = {\apjl},
         year = 2010,
        month = jun,
       volume = {716},
       number = {1},
        pages = {L19-L22},
          doi = {10.1088/2041-8205/716/1/L19},
archivePrefix = {arXiv},
       eprint = {1004.5173},
 primaryClass = {astro-ph.SR},
       adsurl = {https://ui.adsabs.harvard.edu/abs/2010ApJ...716L..19M}
}

\end{document}